# Electro-optic measurement of carrier mobility in an organic thin-film transistor

E.G. Bittle and J.W. Brill

*Department of Physics and Astronomy, University of Kentucky, Lexington, KY 40506-0055, USA*

J.E. Anthony

*Department of Chemistry, University of Kentucky, Lexington, KY 40506-0055, USA*

**Abstract**

We have used an electro-optic technique to measure the position-dependent infrared absorption of holes injected into a thin crystal of the organic semiconductor, 6,13-bis(triisopropylsilylethynyl)-pentacene incorporated in a field-effect transistor. By applying square-wave voltages of variable frequency to the gate or drain, one can measure the time it takes for charges to accumulate on the surface, and therefore determine their mobility.

In recent years, a number of polymer and small molecule organic semiconductors have been suggested for application in thin-film transistors (TFTs).[1,2] A common difficulty has been determining the intrinsic mobility of carriers in these films; for example, apparent mobilities determined from transport measurements are often compromised by barrier effects at metallic contacts.[1] In this paper, we demonstrate that dynamic infrared electro-optic (EO) measurements can be used as an unusual "time-of-flight" technique to measure the mobilities of carriers in field-effect devices.

Previously, Li *et al* have shown that charging a TFT can lead to changes in the IR properties, either through hardening of phonons in the insulating dielectric film[3] or through Raman modes in a semiconducting polymer becoming infrared-active due to polaronic distortions.[4] These static measurements were used to determine the suitability of different dielectrics for organic TFTs. However, since application of a gate voltage pulls charge into the surface layer of the semiconductor and therefore increases the infrared absorption, dynamic EO measurements can be used to measure the time needed for charge to flow and hence the carrier mobility.

For our measurements, we use bottom contact TFTs with thin (~ 40 nm) crystals of p-type 6,13-bis(triisopropylsilylethynyl)-pentacene (TIPS-Pn)[2,5,6] as the active semiconductor on a ~ 1 μm thick dielectric film of the cross-linked polymer poly(vinylphenol) (PVP)[7] cast on a gold gate electrode. The source and drain electrodes are ~30 nm thick evaporated gold films, each ~ 2mm wide and separated by an L = 400 μm long channel. TIPS-Pn crystals were grown from toluene solution between the electrodes, as shown in Figure 1. The molecules self assemble with the TIPS sidegroups adjacent to the substrate so that the pentacene backbones form canted stacks along the crystal **a**-axis, roughly parallel to the channel.[2,6] For the TFT measured, five crystals, between 150 and 500 mm wide, spanned the channel between the source and drain. The IV characteristics of this TFT are shown in the Figure 2 (inset). For $V_{gs}$ = -50 V, 0.4% of the molecules on the surface are charged (using a dielectric capacitance value of 2 nF/cm$^2$), giving saturation and linear hole mobilities of $\mu_{sat} = 0.14$ and $\mu_{lin} = 0.06$ cm$^2$/V·s.

Our EO measurements were made on a ~ 200 μm wide crystal. The (unpolarized) mid-IR reflectance of the sample, mostly determined by reflectance from the gold gate with absorption features from both the TIPS-Pn and PVP films is shown in Figure 2. In this spectral region, there are two absorption features due to TIPS-Pn molecular vibrations and one due to the polymer film, as indicated.

For our EO measurements,[8] chopped infrared light from a tunable diode laser was focused on to a ~ 25 μm long spot on the crystal using an infrared microscope (see Figure 1). The light was polarized perpendicular to the channel (~52$^o$ from the pentacene axes). A unipolar square-wave (SW) voltage was applied to either the gate or drain and the relative change in reflectance, $\Delta R/R$, found by taking the ratio of the signals at the SW ($\omega$) and chopping frequencies. (In general, the responses both in-phase and in quadrature with the SW were measured, but the quadrature response was relatively small at the low frequencies of Figures 1,2.) Because of mechanical and electronic noise from the laser cooling system and detector, respectively, the signal/noise ratio of the EO signal became very poor for $\omega/2\pi < 100$ Hz.

The electro-reflectance spectrum at a point in the center of the channel, measured with a -50 V SW applied to the gate and $V_{ds} = 0$, is shown in Figure 2. The negative values for $\Delta R$ indicate that the reflectance decreases when voltage is applied to the gate. The SW voltage dependence measured at 907 $cm^{-1}$ is shown in Figure 3 (inset); the linear dependence on $V_{gs}$ suggest that the decrease in reflectance is due to the absorption by the gate induced TIPS-Pn surface charges. We approximate $\Delta R/R \sim 2\alpha_{2D}$, where $\alpha_{2D} = \rho_{2D}\, \mu_{lin}/\varepsilon_0 c$ is the fraction of energy absorbed in each pass through the surface charge density $\rho_{2D}$, giving $\Delta R/R \sim 5 \times 10^{-6}$, about an order of magnitude less than our measured value but reasonably good agreement in view of the approximations made. In particular, this simple model does not account for the observed increase in $|\Delta R/R|$ with optical frequency.

It is also surprising that there are no spectral changes, out of the noise, in the EO signal at the TIPS-Pn phonon frequencies, as the phonon frequencies are expected to be sensitive to the molecular charge; e.g. for pentacene, changes of > 20 $cm^{-1}$ have been observed.[9] From the magnitude and width of the reflectance spectral features, and assuming that the crystal is ~ 25 molecules thick, our noise level ($\Delta R/R < 10^{-5}$) suggests that the phonon frequencies change less than 4 $cm^{-1}$ with charging. Alternatively, the lack of spectral anomalies may be an indication that charges in the surface layer hop between molecules in times less than the phonon periods.

We now discuss the position and frequency dependence of the EO response for the case ($V_{ds}$ = 0 and $V_{gs}$ = -50 V SW) when there is only "induced current" and the case ($V_{gs}$ = -50 V dc and $V_{ds}$ = -50 V SW) when there is also "applied current". As shown in Figure 3, the EO signal with "induced current" (at low SW frequency, 505 Hz ) is independent of position in the channel, as expected for uniform charging of the crystal surface. For "applied current" and low frequency, no oscillating signal is observed adjacent to the source while the full EO response is observed adjacent to the drain (with phase opposite to the “no-current” case), as expected. Now, $\Delta R/R$ varies linearly with position between the two electrodes, implying that the EO response (i.e. the local surface charge) only depends on the “vertical” electric field; we cannot resolve (within our spatial resolution and noise level) any possible non-linearities near the electrodes due to charge injection.[4]

The frequency dependence for the “induced current” case, at a point in the center of the channel and at a point adjacent to the drain, is shown in Figure 4; both the quadrature and in-phase responses are shown. In the center of the channel, the response appears relaxational, i.e. $\Delta R/R = \Delta R/R_0 / (1 - i\omega\tau)$, with $1/2\pi\tau = (1.0 \pm 0.2)$ kHz. The response adjacent to the contacts is more than an order of magnitude faster and not simply relaxational, probably reflecting our finite spatial resolution. The slow response in the center reflects the time needed for charges to reach (leave) the center of the sample when the gate voltage is applied (removed). Since during charging or discharging the average longitudinal electric field $\langle E \rangle \sim (V_{gs}/2)/(L/2)$, we expect $\tau \sim (L/2)/(\mu\langle E \rangle) \sim L^2/(2\mu V_{gs})$, giving a value of $\mu \sim (0.10 \pm 0.02)$ $cm^2/V{\cdot}s$, consistent with mobility estimates from the IV curves and showing that, for this crystal, the “transport mobility” is really determined by scattering in the crystal and not due to contact barriers. We should mention that for many other crystals we’ve measured, even those that appear to have good dc transport properties from IV measurements, no EO response (out of the noise) is observed, even adjacent to the contacts, presumably because contact barriers slow the response to frequencies < 100 Hz.

Also shown in Figure 4 is the frequency dependence (at the center of the channel) for the “applied current” case. As discussed above, EO response in the center of the sample is now halved and inverted with respect to the “induced current” case. However, as shown by the curves, it has the same time constant; i.e. the charging time is not, within our resolution, decreased by having net drain-to-source current present for half cycle. To improve the precision of our mobility estimates and to observe changes in the dynamics (or spatial dependence) of surface charging due to current flow will require us to improve the signal/noise ratio of our measurements, e.g. by increasing our signals by using thinner (or higher dielectric constant) insulating films.[10]

In conclusion, we have shown that infrared EO measurements can be used to estimate the mobilities of field induced charges in TFTs, and so may be useful in cases where the mobility determined by transport measurements is ambiguous, e.g. because the induced surface charge density is uncertain. EO measurements may also be useful in determining the extent to which contact barriers hinder transport.

We thank K.-W. Ng and J.P. Straley for helpful discussions. This work was supported by the National Science Foundation, grants # DMR-0800367, CHE- 0749473, and EPS-0814194.

[1] I.H. Campbell and D.L. Smith, Solid State Physics **55**, 1 (2001).

[2]J.E. Anthony, Chem. Rev. **106** (5028) 2006.

[3] Z.Q. Li, G.M. Wang, K.J. Mikolaitis, D. Moses, A.J. Heeger, and D.N. Basov, App. Phys. Lett.**86**, 223506 (2005).

[4] Z.Q. Li, G.M. Wang, BN. Sai, D. Moses, M.C. Martin, M. Di Ventra, A.J. Heeger, and D.N. Basov, Nano Lett. **6**, 224 (2006).

5 O. Ostroverkhova, D.G. Cooke, F.A. Hegmann, R.R. Tykwinski, S.R. Parkin, and J.E. Anthony, App. Phys. Lett. **89**, 192113 (2006).

[6] J.-H. Kwon, J.-H. Seo, S.-I. Shin, K.-H. Kim, D.H. Choi, I.B. Kang, and B.-K. Ju, Trans. Elec. Devices **55**, 500 (2008).

[7] S.C. Lim, S.H. Kim, J.B. Koo, J.H. Lee, C.H. Ku, Y.S. Yang, and T. Zyung, App. Phys. Lett. **90**, 173512 (2007).

[8] L. Ladino, J.W. Brill, M. Freamat, M. Uddin, and D. Dominko, Phys. Rev. B **74**, 115104 (2006).

[9] M. Brinkmann, V.S. Videva, A. Bieber, J.J. Andre, P. Turek, L. Zuppiroli, P. Bugnon, M. Schaer, F. Nuesch, and R. Humphrey-Baker, J. Phys. Chem. A **108** (8170) 2004.

[10] Unfortunately, samples we've so far prepared with thinner PVP films had small (<< 10 μm) pinholes which gave very large, overwhelming EO signals which varied quadratically with $V_{gs}$. Presumably, small grains of TIPS-Pn are pulled into the pinholes by the applied field.

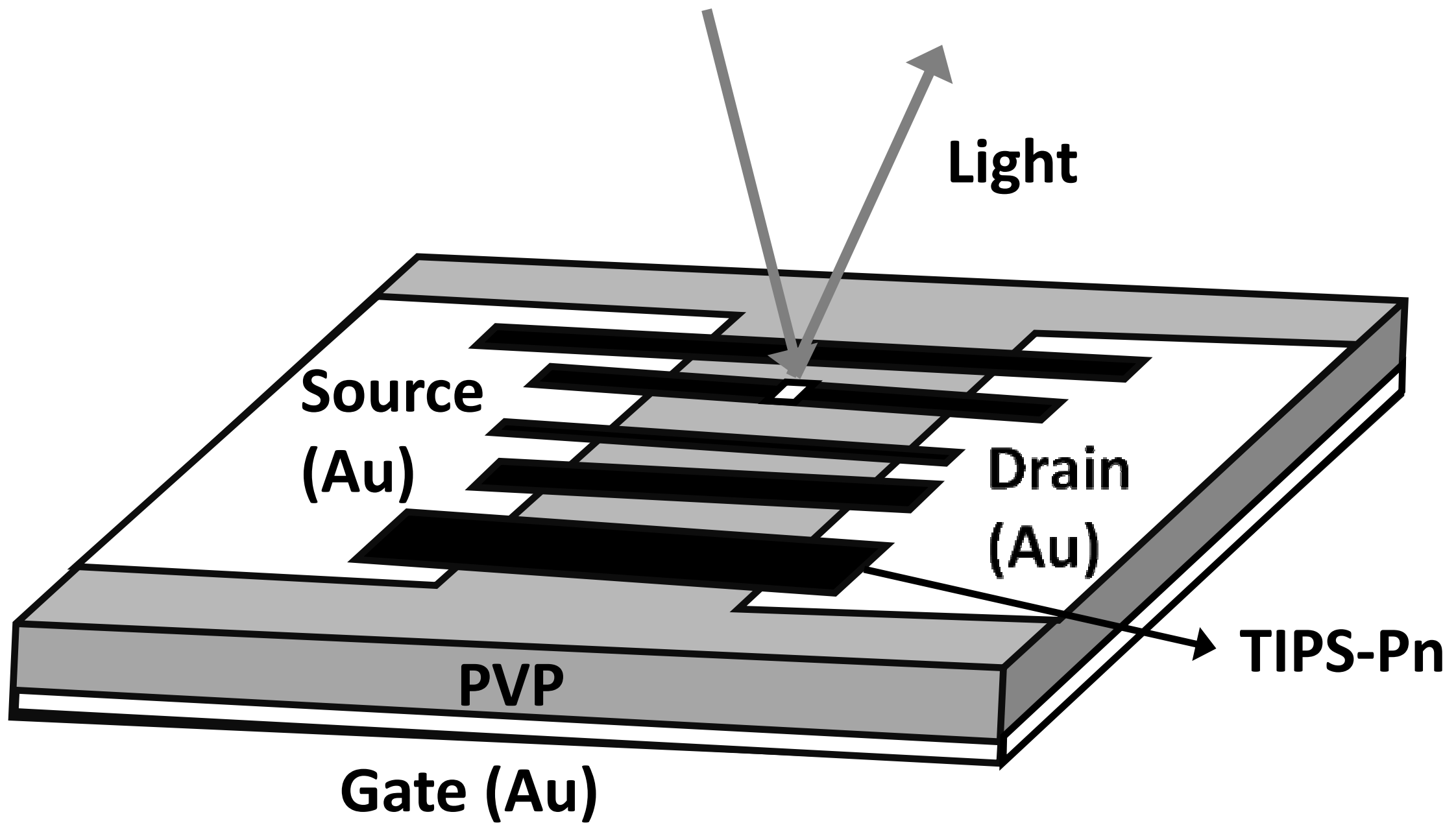

**Figure 1.** Schematic (not to scale) of TIPS-Pn TFT.

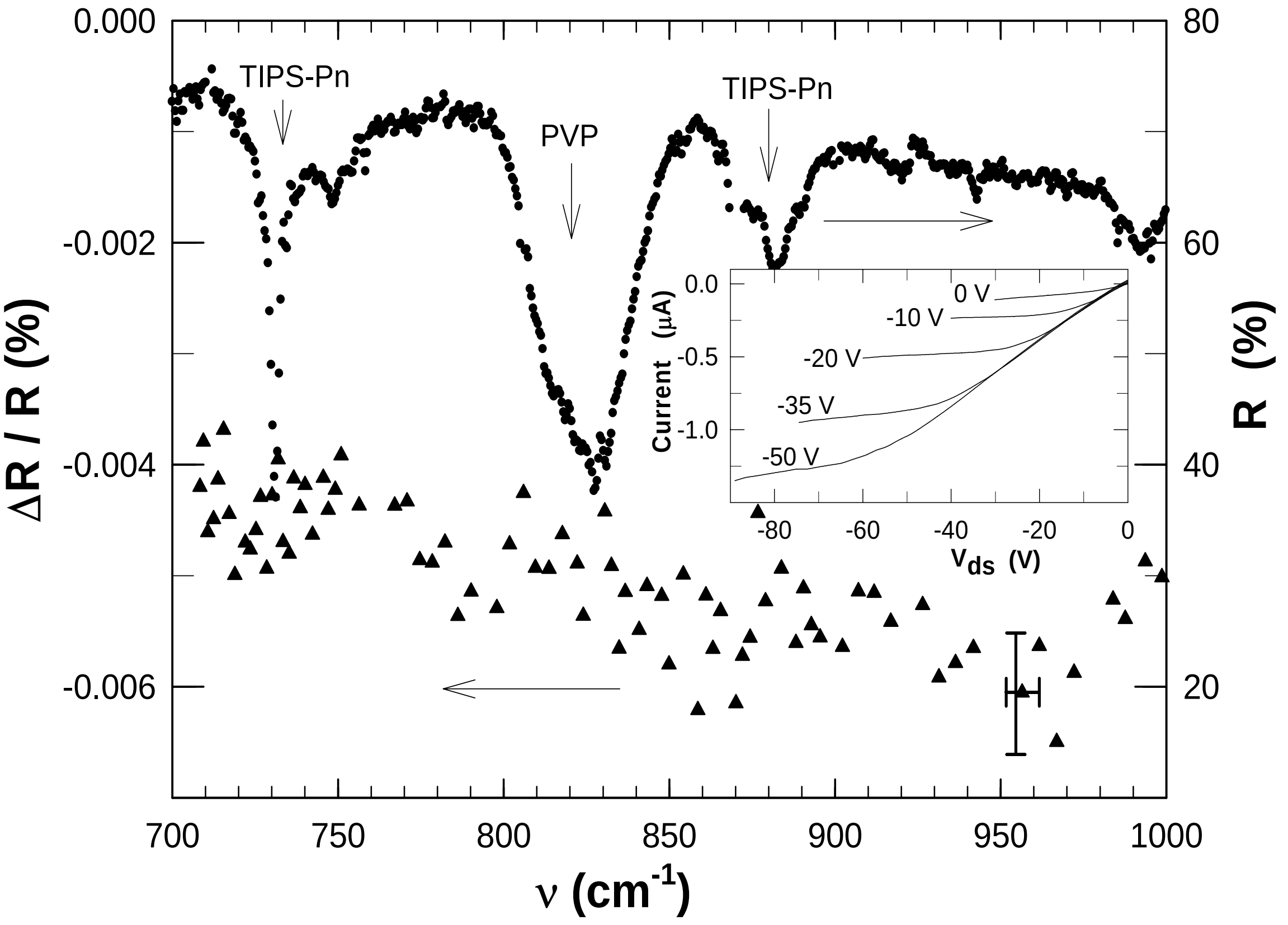

**Figure 2.** Triangles: Electro-reflectance spectrum for a $V_{gs}$ = -50 V square wave ($\omega/2\pi$ = 505 Hz, in-phase response) with $V_{ds}$ =0 (left scale). Circles: Reflectance spectrum (right scale) of the TFT. The laser emits in a few modes giving a typical resolution of ~ 15 $cm^{-1}$. Inset: Current vs. $V_{ds}$ for different vales of (dc) $V_{gs}$.

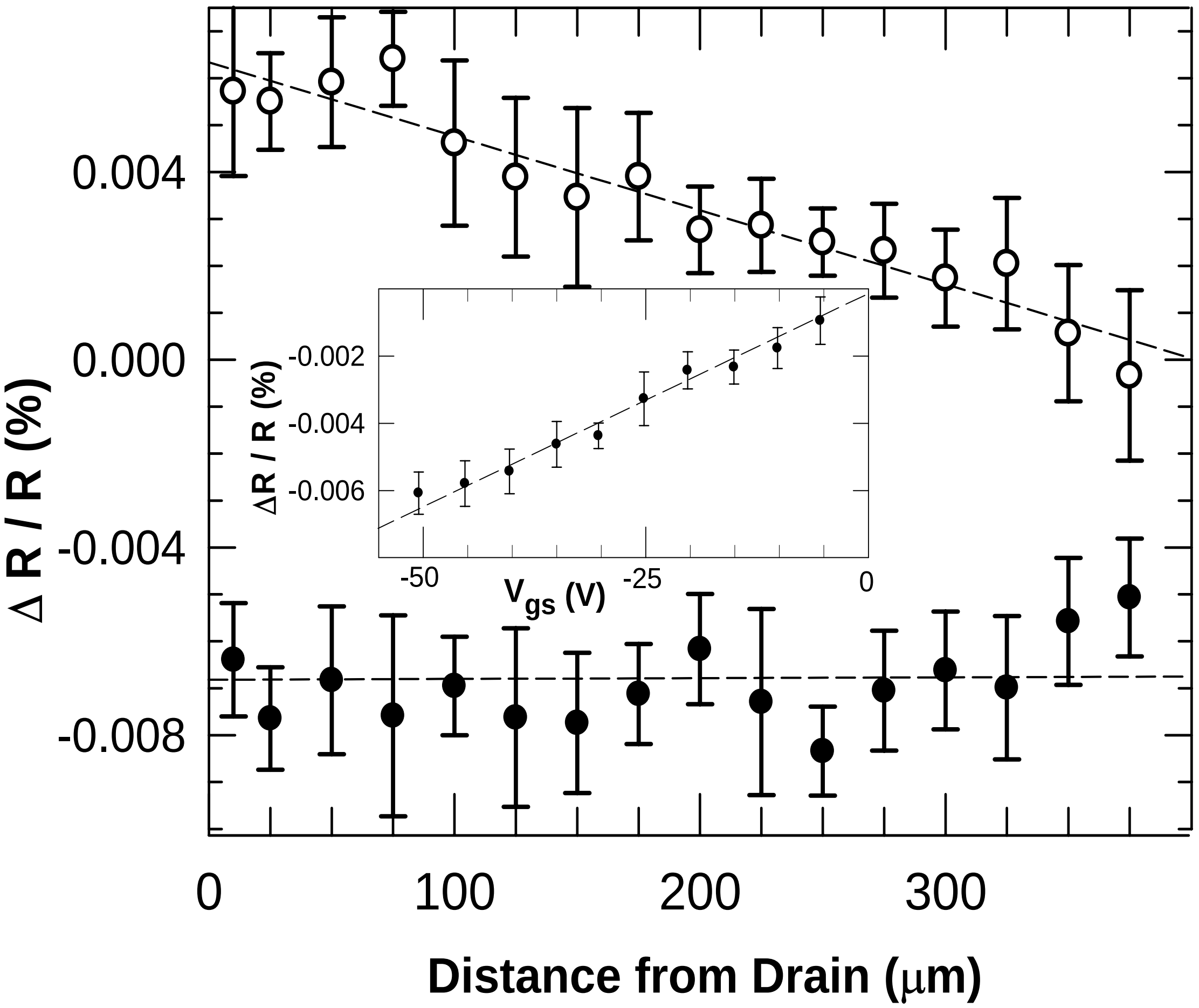

**Figure 3.** ΔR/R measured (at ν = 907 $cm^{-1}$) in-phase with the applied (505 Hz) square-wave as a function of position in the channel. Solid circles: "induced current". Open circles: "applied current". Inset: ΔR/R vs. the amplitude of the 505 Hz $V_{gs}$ square-wave, with $V_{ds}$=0, at the center of the channel. (The lines are guides to the eye.)

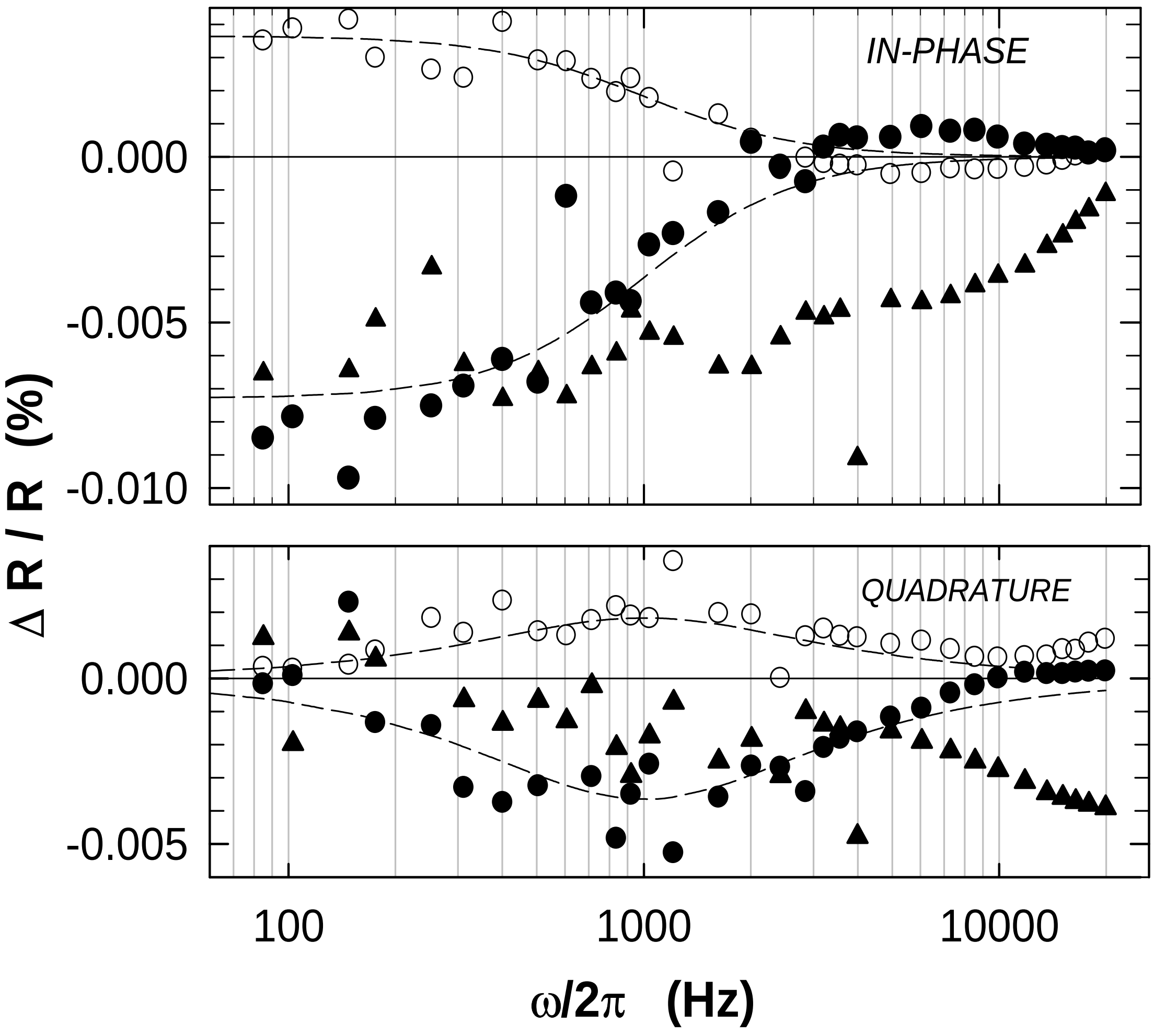

**Figure 4.** Dependence of ΔR/R (in-phase and in quadrature with the applied square wave) vs. SW frequency. Solid circles: center of channel, "induced current". Solid triangles: adjacent to drain, "induced current". Open circles: "applied current", center of channel. The dashed curves show relaxational behavior, with $1/2\pi\tau$ = 1 kHz, for the center of channel cases.